\documentstyle[epsfig,rotating,float,aps,preprint,local]{revtex}
\tightenlines
\begin{document}
\setcounter{figure}{0}

\title{\bf Dark Energy in Light of the Cosmic Horizon}
\author{\bf Fulvio Melia\footnote{Sir Thomas Lyle Fellow and Miegunyah Fellow.}}
\address{Department of Physics and Steward Observatory, \\
The University of Arizona,\
Tucson, Arizona 85721, USA\ }
\maketitle
\def\ref{\par\vskip 12pt \noindent \hangafter=1 \hangindent=22.76pt}
\begin{abstract}
Based on dramatic observations of the cosmic microwave background
radiation with WMAP and of Type Ia supernovae with the Hubble
Space Telescope and ground-based facilities, it is now generally
believed that the Universe's expansion is accelerating. Within
the context of standard cosmology, this type of evolution leads
to the supposition that the Universe must contain a third `dark'
component of energy, beyond matter and radiation. However, the
current data are still deemed insufficient to distinguish between
an evolving dark energy component and the simplest model of a
time-independent cosmological constant. In this paper, we examine
the role played by our cosmic horizon $R_0$ in our interrogation of
the data, and reach the rather firm conclusion that the existence
of a cosmological constant is untenable. The observations are
telling us that $R_0\approx ct_0$, where $t_0$ is the perceived
current age of the Universe, yet a cosmological constant would
drive $R_0$ towards $ct$ (where $t$ is the cosmic time) only once,
and that would have to occur right now. In contrast, scaling
solutions simultaneously eliminate several conundrums in the
standard model, including the `coincidence' and `flatness'
problems, and account very well for the fact that $R_0\approx 
ct_0$. We show in this paper that for such dynamical dark
energy models, either $R_0=ct$ for all time (thus eliminating
the apparent coincidence altogether), or that what we believe
to be the current age of the universe is actually the horizon
time $t_h\equiv R_0/c$, which is always shorter than $t_0$.
Our best fit to the Type Ia supernova data indicates that
$t_0$ would then have to be $\approx 16.9$ billion years.
Though surprising at first, an older universe such as this
would actually eliminate several other long-standing problems
in cosmology, including the (too) early appearance of supermassive
black holes (at a redshift $>6$) and the glaring deficit of dwarf
halos in the local group.
\end{abstract}
\vskip 0.3in

\newpage
\section{Introduction}
Over the past decade, Type Ia supernovae have been used successfully as standard
candles to facilitate the acquisition of several important cosmological parameters.
On the basis of this work, it is now widely believed that the Universe's expansion
is accelerating (Riess et al. 1998; Perlmutter et al. 1999). In standard cosmology,
built on the assumption of spatial homogeneity and isotropy, such an expansion
requires the existence of a third form of energy, beyond the basic admixture of
(visible and dark) matter and radiation.

One may see this directly from the (cosmological) Friedman-Robertson-Walker (FRW)
differential equations of motion, usually written as
\begin{equation}
H^2\equiv\left(\frac{\dot a}{a}\right)^2=\frac{8\pi G}{3c^2}\rho-\frac{kc^2}{a^2}\;,
\end{equation}
\begin{equation}
\frac{\ddot a}{a}=-\frac{4\pi G}{3c^2}(\rho+3p)\;,
\end{equation}
\begin{equation}
\dot\rho=-3H(\rho+p)\;,
\end{equation}
in which an overdot denotes a derivative with respect to cosmic time $t$, and $\rho$
and $p$ represent, respectively, the total energy density and total pressure.
In these expressions, $a(t)$ is the expansion factor, and $(r,\theta,\phi)$ are
the coordinates in the comoving frame. The constant $k$ is $+1$ for a closed
universe, $0$ for a flat (or open) universe, and $-1$ for an open universe.

Following convention, we write the equation of state as $p=\omega\rho$.
A quick inspection of equation (2) shows that an accelerated expansion
($\ddot a>0$) requires $\omega<-1/3$. Thus, neither radiation ($\rho_r$, with
$\omega_r=1/3$), nor (visible and dark) matter ($\rho_m$, with $\omega_m\approx 0$)
can satisfy this condition, leading to the supposition that a third `dark'
component $\rho_d$ (with $\omega_d<-1/3$) of the energy density $\rho$ must
be present. In principle, each of these contributions to $\rho$ may evolve
according to its own dependence on $a(t)$.

Over the past few years, complementary measurements (Spergel et al. 2003) of the
cosmic microwave background (CMB)
radiation have indicated that the Universe is flat (i.e., $k=0$), so $\rho$ is at
(or very near) the ``critical" density $\rho_c\equiv 3c^2H^2/8\pi G$. But among the
many peculiarities of the standard model is the inference, based on current
observations, that $\rho_d$ must
itself be of order $\rho_c$. Dark energy is often thought to be the manifestation
of a cosmological constant, $\Lambda$, though no reasonable explanation has yet been
offered as to why such a fixed, universal density ought to exist at this scale. It is
well known that if $\Lambda$ is associated with the energy of the vacuum in quantum
theory, it should have a scale representative of phase transitions in the early
Universe---many, many orders of magnitude larger than $\rho_c$.

Many authors have attempted to circumvent these difficulties by proposing
alternative forms of dark energy, including Quintessence (Wetterich 1988;
Ratra and Peebles 1988), which represents an evolving canonical scalar
field with an inflation-inducing potential, a Chameleon field (Khoury and
Weltman 2004; Brax et al. 2004) in which the scalar field couples to the
baryon energy density and varies from solar system to cosmological scales,
and modified gravity, arising out of both string motivated (Davli
et al. 2000), or General Relativity modified (Capozziello et al. 2003;
Carroll et al. 2004) actions, which introduce large length scale corrections
modifying the late time evolution of the Universe. The actual number of
suggested remedies is far greater than this small, illustrative sample.

Nonetheless, though many in the cosmology community suspect that some sort
of dynamics is responsible for the appearance of dark energy, until now the
sensitivity of current observations has been deemed insufficient to distinguish
between an evolving dark energy component and the simplest model of a
time-independent cosmological constant $\Lambda$ (see, e.g., Corasaniti et al.
2004). This conclusion, however, appears to be premature, given that the
role of our cosmic horizon has not yet been fully folded into the
interrogation of current observations.

The purpose of this paper is to demonstrate that a closer scrutiny of the
available data, if proven to be reliable, can in fact already delineate
between evolving and constant dark energy theories, and that a simple
cosmological constant $\Lambda$, characterized by a fixed $\omega_d=
\omega_\Lambda=-1$, is almost certainly ruled out. In \S~2 of this paper, we will
introduce the cosmic horizon and discuss its evolution in time, demonstrating
how measurements of the Hubble constant $H$ may be used to provide strict
constraints on $\omega$, independent of Type Ia supernova data. We will
compare these results with predictions of the various classes of dark
energy models in \S\S~3 and 4, and conclude with a discussion of the consequences
of these comparisons in \S~5.

\section{The Cosmic Horizon}
The Hubble Space Telescope Key Project on the extragalactic distance
scale has measured the Hubble constant $H$ with unprecedented accuracy (Mould et
al. 2000), yielding a current value $H_0\equiv H(t_0)=71\pm6$ km s$^{-1}$ Mpc$^{-1}$.
(For $H$ and $t$, we will use subscript ``0" to denote cosmological values pertaining
to the current epoch.) With this $H_0$, we infer that $\rho(t_0)=\rho_c\approx 
9\times 10^{-9}$ ergs cm$^{-3}$.

Given such precision, it is now possible to accurately calculate the
radius of our cosmic horizon, $R_0$, defined by the condition
\begin{equation}
\frac{2GM(R_0)}{c^2}=R_0\;,
\end{equation}
where $M(R_0)=(4\pi/3) R_0^3\rho/c^2$. In terms of $\rho$,
$R_0=({3c^4/8\pi G\rho})^{1/2}$ or, more simply, $R_0=c/H_0$ in a flat universe.
This is the radius at which a sphere encloses sufficient mass-energy to turn it
into a Schwarzschild surface for an observer at the origin of the coordinates
$(cT,R,\theta,\phi)$, where $R=a(t)r$, and $T$ is the time corresponding to
$R$ (Melia 2007).

When the Robertson-Walker metric is written in terms of these observer-dependent
coordinates, the role of $R_0$ is to alter the intervals of time we measure
(using the clocks fixed to our origin), in response to the increasing
spacetime curvature induced by the mass-energy enclosed by a sphere with
radius $R$ as $R\rightarrow R_0$. The time $t$ is identical to
$T(R)$ only at the origin ($R=0$). For all other radii, our measurement of a
time interval $dT$ necessarily comes with a gravitational time dilation which
diverges when $R\rightarrow R_0$. It is therefore physically impossible for
us to see any process occurring beyond $R_0$, and this is why the recent
observations have a profound impact on our view of the cosmos. For example,
from the Hubble measurement of $\rho(t_0)$, we infer that $R_0\approx 13.5$ billion
light-years; this is the maximum distance out to which measurements of
the cosmic parameters may be made at the present time.

Let us now consider how this radius evolves with the universal expansion.
Clearly, in a de Sitter universe with a constant $\rho$, $R_0$ is fixed
forever. But for any universe with $\omega\not=-1$, $R_0$ must be a function
of time. From the definition of $R_0$ and equation (3), we see that
\begin{equation}
{\dot R}_0=\frac{3}{2}(1+\omega)c\;,
\end{equation}
a remarkably simple expression that nonetheless leads to several important
conclusions regarding our cosmological measurements (Melia 2008).
We will use it here to distinguish between constant and evolving dark
energy theories.

Take $t$ to be some time in the distant past (so that $t\ll t_0$). Then,
integrating equation (5) from $t$ to $t_0$, we find that
\begin{equation}
R_0(t_0)-R_0(t)=\frac{3}{2}(1+\langle\omega\rangle)ct_0\;,
\end{equation}
where
\begin{equation}
\langle\omega\rangle\equiv \frac{1}{t_0}\int_{t}^{t_0}\omega\,dt
\end{equation}
is the time-averaged value of $\omega$ from $t$ to the present time.

Now, for any $\langle\omega\rangle>-1$, $\rho$ drops as the universe
expands (i.e., as $a(t)$ increases with time), and since
$R_0\sim\rho^{-1/2}$, clearly $R_0(t)\ll R_0(t_0)$. Therefore,
\begin{equation}
R(t_0)\approx\frac{3}{2}(1+\langle\omega\rangle)ct_0\;.
\end{equation}
The reason we can use the behavior of $R_0$ as the universe expands
to probe the nature of dark energy is that the latter directly impacts
the value of $\langle\omega\rangle$. A consideration of how the cosmic
horizon $R_0$ evolves with time can therefore reveal whether or not
dark energy is dynamically generated. Indeed, we shall see in the next
section that the current observations, together with equation (8), are
already quite sufficient for us to differentiate between the various models.

Before we do that, however, we can already see from this expression
why the standard model of cosmology contains a glaring inconsistency
(Melia 2008). From WMAP observations (Spergel et al. 2003), we infer that
the age $t_0$ of the universe is $\approx 13.7$ billion years. Since
$R_0\approx 13.5$ billion light-years, this can only occur if
$\langle\omega\rangle\le -1/3$. Of course, this means that the existence
of dark energy (with such an equation of state) is required by the
WMAP and Hubble observations alone, independently of the Type Ia
supernova data. But an analysis of the latter (see below) reveals that
the value $\langle\omega\rangle=-1/3$ is ruled out, so in fact
$\langle\omega\rangle<-1/3$. That means that $R_0\not=ct_0$; in fact,
$R_0$ must be less than $ct_0$, which in turns suggests that the
universe is older than we think. What we infer to be the time
since the Big Bang, is instead the ``horizon" time $t_h\equiv
R_0/c$, which must be shorter than $t_0$. As discussed in Melia
(2008), this has some important consequences that may resolve
several long-standing conflicts in cosmology. Through our analysis
in the next section, we will gain a better understanding of this
phenomenon, which will permit us to calculate $t_0$ more precisely.

\begin{figure}
\center{\includegraphics[scale=0.5,angle=-90]{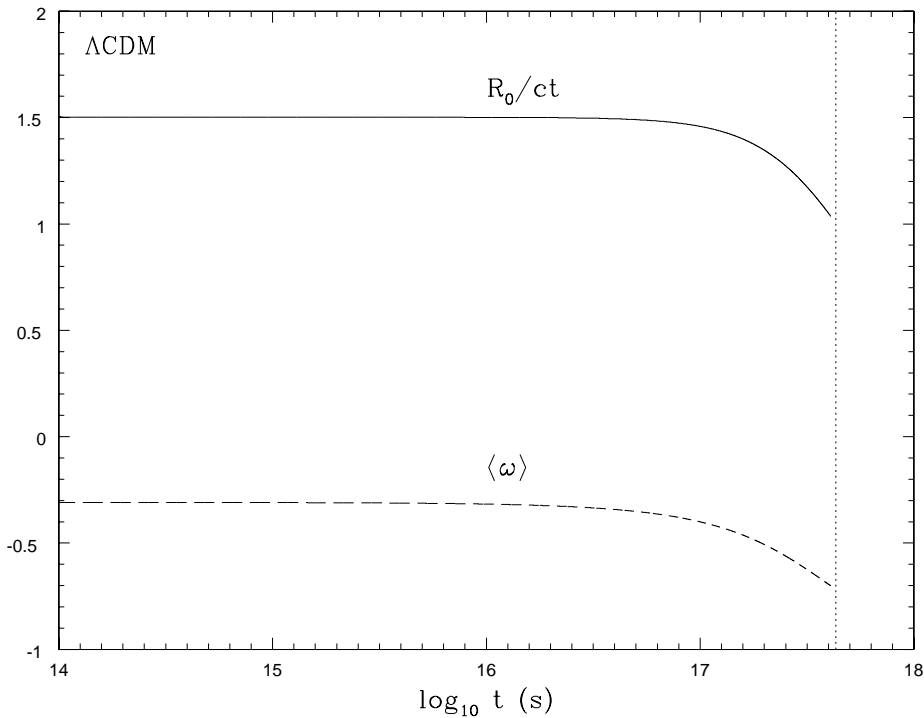}
\caption{Plot of the horizon radius $R_0$ in units of $ct$, and
$\langle\omega\rangle$, the equation of state parameter $\omega\equiv p/\rho$
averaged over time from $t$ to $t_0$. The asymptotic value of $\langle\omega
\rangle$ for $t\rightarrow 0$ is approximately $-0.31$. These results are from
a calculation of the universe's expansion in a $\Lambda$CDM cosmology, with
matter energy density $\rho_m(t_0)=0.3\rho_c(t_0)$ and a cosmological constant
$\rho_d=\rho_\Lambda=0.7\rho_c(t_0)$, with $\omega_d\equiv\omega_\Lambda=-1$.}}
\end{figure}

\section{The Cosmological Constant}
The standard model of cosmology contains a mixture of cold dark matter
and a cosmological constant with an energy density fixed at the
current value, $\rho_d(t)\equiv\rho_\Lambda(t)\approx 0.7\,\rho_c(t_0)$,
and an equation of state with $\omega_d\equiv\omega_\Lambda=-1$. Known
as $\Lambda$CDM, this model has been reasonably successful in
accounting for large scale structure, the cosmic microwave background
fluctuations, and several other observed cosmological properties (see,
e.g., Ostriker and Steinhardt 1995; Spergel et al. 2003; Melchiorri
et al. 2003).

But let us now see whether $\Lambda$CDM is also consistent with
our understanding of $R_0$. Putting $\rho=\rho_m(t)+\rho_\Lambda$,
where $\rho_m$ is the time-dependent matter energy density, we may
integrate equation (5) for a $\Lambda$CDM cosmology, starting at the
present time $t_0$, and going backwards towards the era when radiation
dominated $\rho$ (somewhere around 100,000 years after the Big
Bang). Figure~1 shows the run of $R_0/ct$ as a function of time, along
with the time-averaged $\omega$ given in equation (7). The present epoch
is indicated by a vertical dotted line. The calculation begins at the
present time $t_0$, with the initial value $R_0=(3/2)(1+
\langle\omega\rangle_\infty)ct_0$, where $\langle\omega\rangle_\infty$
($\approx -0.31$) is the equation-of-state parameter $\omega$
averaged over the entire universal expansion, obtained by an
iterative convergence of the solution to equation (5). In $\Lambda$CDM,
the matter density increases towards the Big Bang, but $\rho_\Lambda$
is constant, so the impact of $\omega_d$ on the solution vanishes as
$t\rightarrow 0$ (see the dashed curve in figure~1). Thus, as expected,
$R_0/ct\rightarrow 3/2$ early in the Universe's development.

What is rather striking about this result is that in $\Lambda$CDM,
$R_0(t)$ approaches $ct$ only once in the entire history of the
Universe---and this is only because we have imposed this requirement
as an initial condition on our solution.  There are many peculiarities
in the standard model, some of which we will encounter shortly, but the
unrealistic coincidence that $R_0$ should approach $ct_0$ only at the
present moment must certainly rank at---or near---the top of this list.

\vskip -0.1in
\begin{figure}
\center{\includegraphics[scale=0.5,angle=-90]{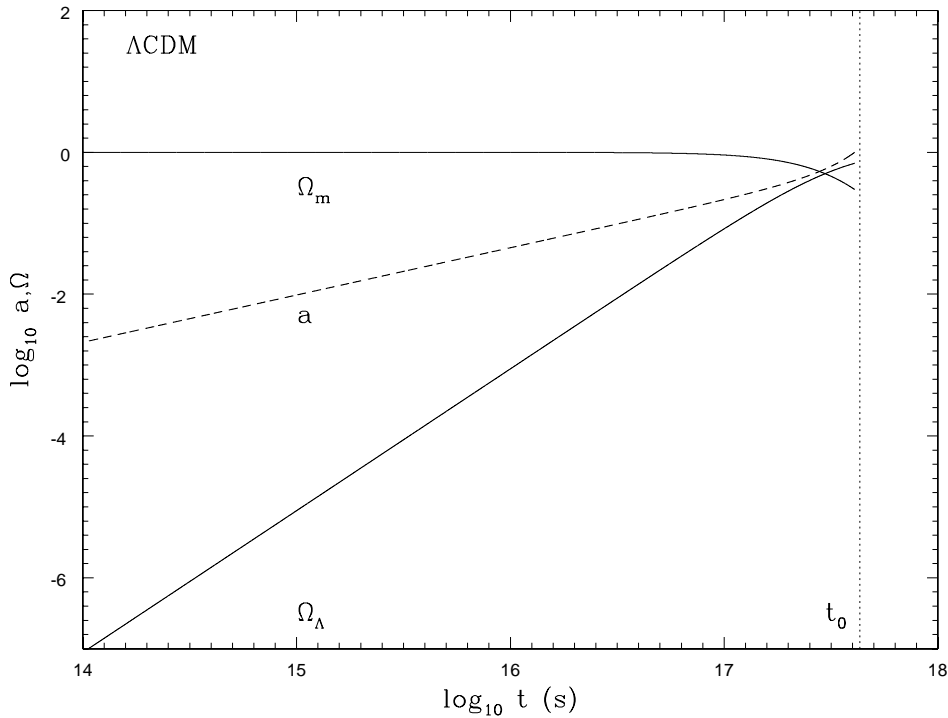}
\caption{Plot of the matter energy density parameter
$\Omega_m\equiv\rho_m(t)/\rho_c(t)$, the cosmological constant energy
density parameter $\Omega_\Lambda\equiv\rho_\Lambda(t)/\rho_c(t)$,
and the expansion factor $a(t)$, as functions of cosmic time $t$ for
the same $\Lambda$CDM cosmology as that shown in figure~1. Since the
energy density associated with $\Lambda$ is constant, $\Omega_m
\rightarrow 1$ as $t\rightarrow 0$. A notable (and well-known) peculiarity of
this cosmology is the co-called ``coincidence problem," so dubbed because
$\Omega_\Lambda$ is approximately equal to $\Omega_m$ only in the current
epoch.}}
\end{figure}
\vskip 0.1in

Though it may not be immediately obvious, this seemingly contrived
scenario is actually related to the so-called ``coincidence problem" in
$\Lambda$CDM cosmology, arising from the peculiar near-simultaneous
convergence of $\rho_m$ and $\rho_\Lambda$ towards $\rho_c$ in the
present epoch. Just as $R_0/ct\rightarrow 1$ only at $t_0$ (see
figure~1), so too $\rho_\Lambda/\rho_m\sim 1$ just now (see figure~2).
Even so, it is one thing to suppose that $\rho_m\sim\rho_\Lambda$
only near the present epoch; it takes quite a leap of faith to
go one step further and believe that $R_0\rightarrow ct_0$ only now.
This odd behavior must certainly cast serious doubt on the viability
of $\Lambda$CDM cosmology as the correct description of the Universe.

A confirmation that the basic $\Lambda$CDM model is simply not viable
is provided by attempts (figure~3) to fit the Type Ia supernova data
with the evolutionary profiles shown in figures~1 and 2. The comoving
coordinate distance from some time $t$ in the past to the present is
$\Delta r=\int_t^{t_0} c\,dt/a(t)$. With $k=0$, the luminosity distance
$d_L$ is $(1+z)\Delta r$, where the redshift $z$ is given by $(1+z)=1/a$,
in terms of the expansion factor $a(t)$ plotted in figure~2.

\vskip -0.1in
\begin{figure}
\center{\includegraphics[scale=0.5,angle=-90]{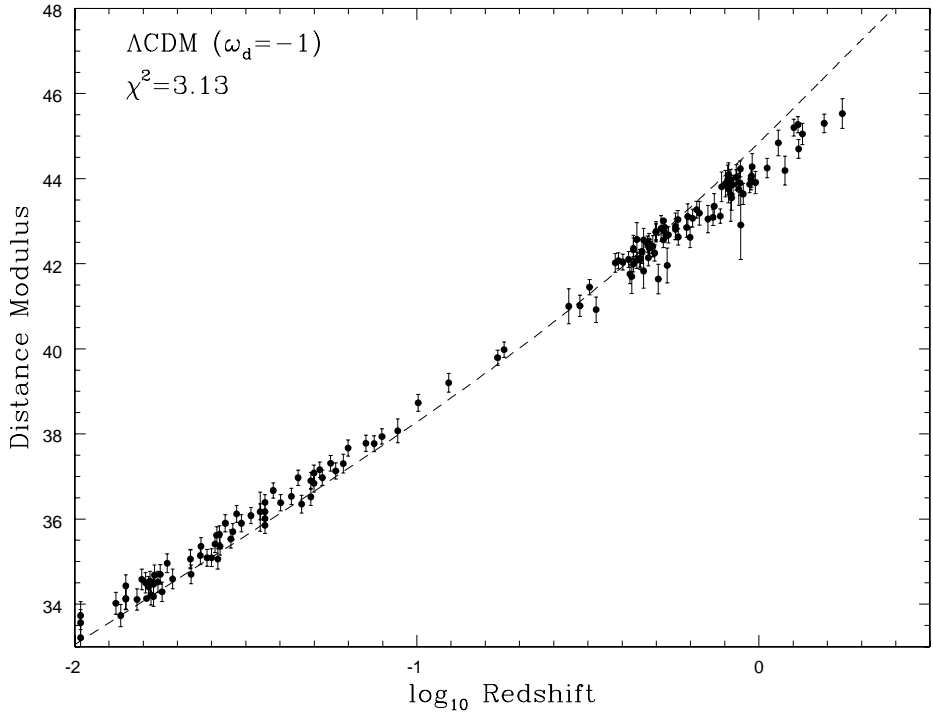}
\caption{Plot of the observed distance modulus versus redshift for
well-measured distant Type Ia supernovae (Riess et al. 2004). The dashed curve
shows the theoretical distribution of magnitude versus redshift for the $\Lambda$CDM
cosmology described in the text and presented in figures~1 and 2. Within the
context of this model, the best fit corresponds to $\Delta=26.57$ and a reduced
$\chi^2\approx 3.13$, with $180-1=179$ d.o.f.}}
\end{figure}
\vskip 0.1in

The data in figure~3 are taken from the ``gold" sample of Riess
et al. (2004), with coverage in redshift from $0$ to $\sim 1.8$.
The distance modulus is $5\log d_L(z)+\Delta$, where $\Delta\approx
25$. The dashed curve in this plot represents the fit based on the
$\Lambda$CDM profile shown in figures~1 and 2, with a Hubble
constant $H_0=71$ km s$^{-1}$ Mpc$^{-1}$ ($\Delta$ is used as
a free optimization parameter in each of figures~3, 6, and 9). The
``best" match corresponds to $\Delta=26.57$, for which the reduced
$\chi^2$ is an unacceptable $3.13$ with $180-1=179$ d.o.f.

Attempts to fix this failure have generally been based on the idea
that there must have been a transition point, at a redshift $\sim
0.5$, from past deceleration to present acceleration (Riess et al.
2004). One can easily see from figure~3 that the 16 high-redshift
Type Ia supernovae, detected at $z>1.25$ with the Hubble Space Telescope,
fall well below the distance expected for them in basic $\Lambda$CDM.
But even if these modifications can somehow be made consistent with
the supernova data, in the end they must still comply with the unyielding
constraints imposed on us by the measured values of $R_0$ and $ct_0$.
No such relief patch can remove the inexplicable (and simply unacceptable)
coincidence implied by the required evolution of $R_0/ct$ towards 1 only
at the present time (figure~1). It is difficult avoiding the conclusion
that $\Lambda$CDM is simply not consistent with the inferred properties
of dark energy in light of the cosmic horizon.

\vskip -0.1in
\begin{figure}
\center{\includegraphics[scale=0.5,angle=-90]{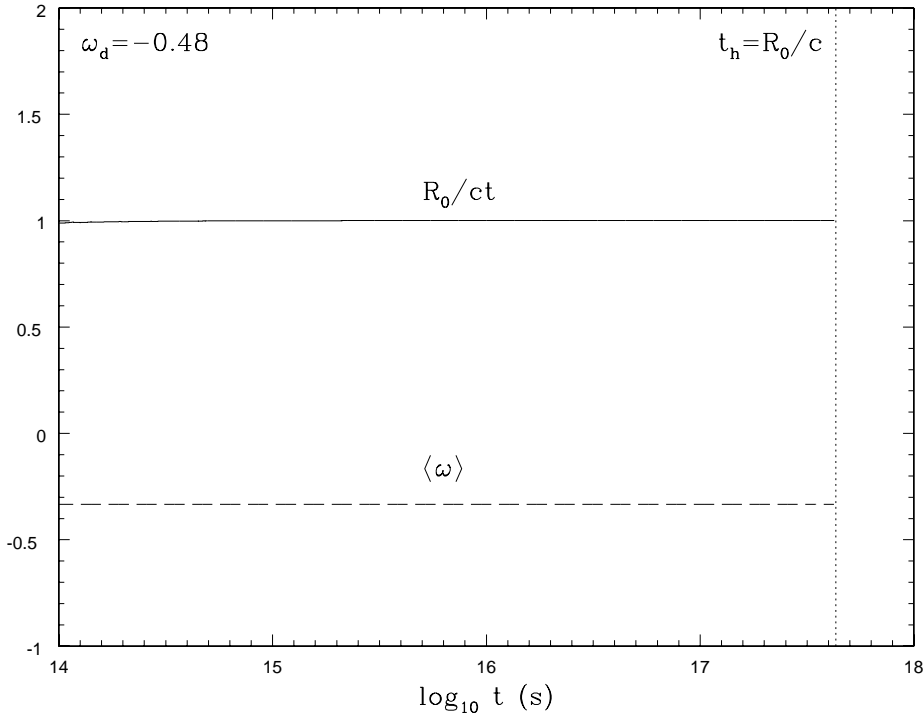}
\caption{Same as figure~1, except for a ``scaling solution"
in which $\rho_d\propto\rho_m$ and $\langle\omega\rangle=-1/3$. The
corresponding dark-energy equation-of-state parameter is $\omega_d\approx
-0.48$. This is sufficiently close to $-0.5$, that it may actually correspond
to this value within the errors associated with the measurement
of $\rho_m(t_0)/\rho_c(t_0)$ and $\rho_d(t_0)/\rho_c(t_0)$. Note that for this
cosmology, $R_0/ct$ is always exactly one. A universe such as this would
do away with the otherwise inexplicable coincidence that $R_0(t_0)=ct_0$
(since it has this value for all $t$), but as we shall see in figure~6, it does
not appear to be consistent with Type Ia supernova data.}}
\end{figure}

\section{Dynamical Dark Energy}
Given the broad range of alternative theories of dark energy that
are still considered to be viable, it is beyond the scope of this paper
to exhaustively study all dynamical scenarios. Instead, we
shall focus on a class of solutions with particular importance to
cosmology---those in which the energy density of the scalar field mimics
the background fluid energy density. Cosmological models in this category
are known as ``scaling solutions," characterized by the relation
\begin{equation}
\frac{\rho_d(t)}{\rho_m(t)}=\frac{\rho_d(t_0)}{\rho_m(t_0)}\approx 2.33
\end{equation}
(some of the early papers on this topic
include the following: Copeland et al. 1998; Liddle and Scherrer 1999;
van den Hoogen et al. 1999; Uzan 1999; de la Macorra and Piccinelli 2000;
Nunes and Mimoso 2000; Ng et al. 2001; Rubano and Barrow 2001).

By far the simplest cosmology we can imagine in this class is that
for which $\omega=-1/3$, corresponding to $\omega_d\approx -1/2$ (within
the errors). This model is so attractive that it almost begs to be
correct. Unfortunately, it is not fully consistent with Type Ia
supernova data, so either our interpretation of current observations
is wrong or the Universe just doesn't work this way. But it's worth
our while spending some time with it because of the shear elegance it
brings to the table.

\vskip -0.1in
\begin{figure}
\center{\includegraphics[scale=0.5,angle=-90]{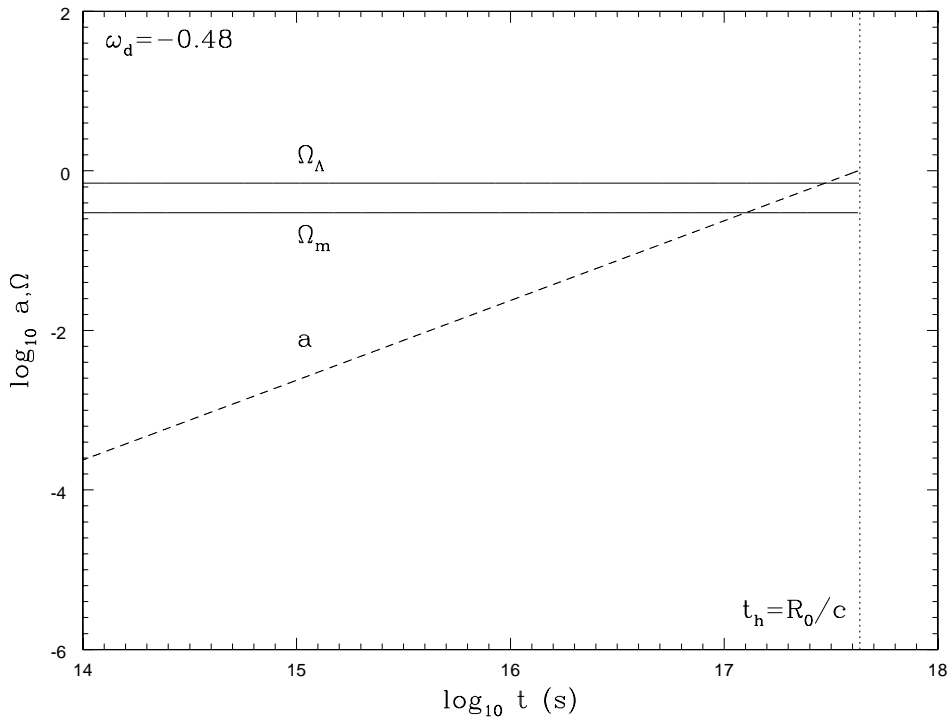}
\caption{Same as figure~2, except for a ``scaling solution"
in which $\rho_d\propto\rho_m$ and $\langle\omega\rangle=-1/3$. The
corresponding dark-energy equation-of-state parameter is $\omega_d\approx
-0.48$.}}
\end{figure}
\vskip 0.1in

To begin with, we see immediately from equation (5) (and illustrated in
figures~4 and 5) that when $\omega=-1/3$, we have $R_0(t)=ct$. Instantly,
this solves three of the major problems in standard cosmology: first, it
explains why $R_0(t_0)$ should be equal to $ct_0$ (because these quantities
are always equal). Second, it completely removes the inexplicable
coincidence that $\rho_d$ and $\rho_m$ should be comparable to each
other only in the present epoch (since they are always comparable to
each other). Third, it does away with the so-called flatness problem
(Melia 2008). To see this, let us return momentarily to equation (1)
and rewrite it as follows:
\begin{equation}
H^2=\left(\frac{c}{R_0}\right)^2\left(1-\frac{kR_0^2}{a^2}\right)\;.
\end{equation}
Whether or not the Universe is asymptotically flat hinges on the
behavior of $R_0/a$ as $t\rightarrow\infty$. But from the definition
of $R_0$ (equation 4), we infer that
\begin{equation}
\frac{d}{dt}\ln R_0=\frac{3}{2}(1+\omega)\frac{d}{dt}\ln a\;.
\end{equation}
Thus, if $\omega=-1/3$,
\begin{equation}
\frac{d}{dt}\ln R_0=\frac{d}{dt}\ln a\;,
\end{equation}
and so
\begin{equation}
H={\rm constant}\times \left(\frac{c}{R_0}\right)\;.
\end{equation}
We also learn from equation (2) that in this universe, $\ddot{a}=0$.
The Universe is coasting, but not because it is empty, as in the Milne
cosmology (Milne 1940), but rather because the change in pressure as it
expands is just right to balance the change in its energy density.

\vskip -0.1in
\begin{figure}
\center{\includegraphics[scale=0.5,angle=-90]{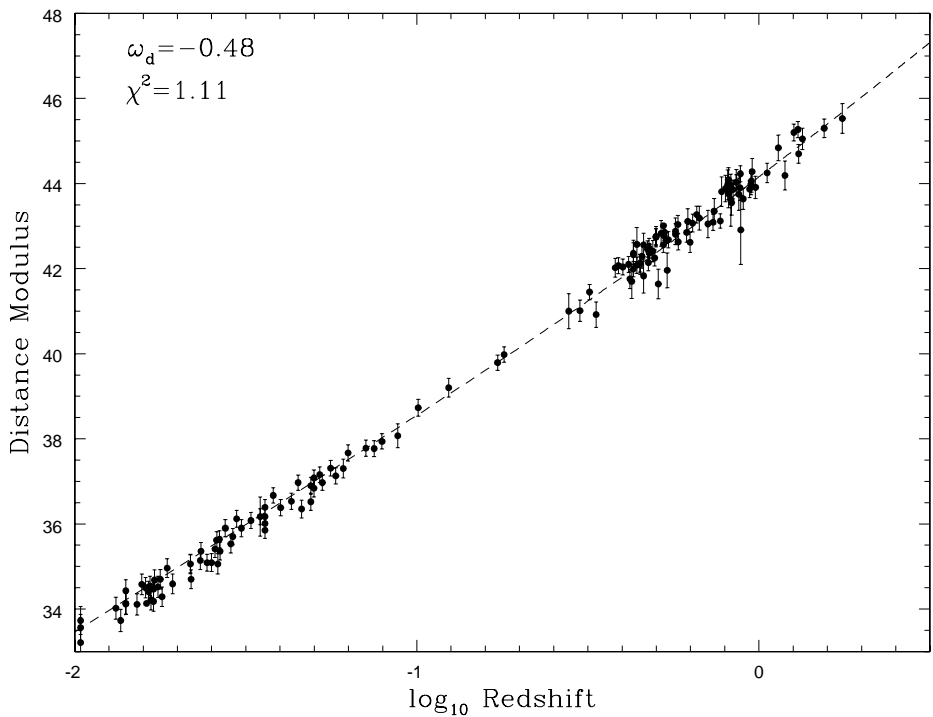}
\caption{Same as figure~3, except for a ``scaling solution"
in which $\rho_d\propto\rho_m$ and $\langle\omega\rangle=-1/3$. The
corresponding dark-energy equation-of-state parameter is $\omega_d\approx
-0.48$. This fit is much better than that corresponding to a $\Lambda$CDM
cosmology (figure~3) but, with a reduced $\chi^2\approx 1.11$ for
$180-1=179$ d.o.f., is still not an adequate representation of
the Type Ia supernova data (Riess et al. 2004).}}
\end{figure}
\vskip 0.1in

All told, these are quite impressive accomplishments for such a simple
model, and yet, it doesn't appear to be fully consistent with Type Ia
supernova data. Repeating the calculation that produced figure~3,
only now for our scaling solution with $\omega=-1/3$, we obtain the
best fit shown in figure~6. This is a significant improvement over
$\Lambda$CDM, but the resultant $\chi^2$ ($\approx 1.11$) is still
unacceptable. Interestingly, if we were to find a slight systematic
error in the distance modulus for the events at $z>1$, which for some
reason has led to a fractional over-estimation in their distance (or,
conversely, a systematic error that has lead to an under-estimation
of the distance modulus for the nearby explosions), the fit would
improve significantly. So our tentative conclusion right now must
be that, although an elegant scaling solution with $\omega=-1/3$
provides a much better description than $\Lambda$CDM of the Universe's
expansion, it is nonetheless still not fully consistent with the
supernova data.

Fortunately, many of the attractive features of an $\omega=-1/3$
cosmology are preserved in the case where $\omega_d=-2/3$,
corresponding to a scaling solution with $\omega\approx-1/2$.
This model fits the supernova data quite strikingly, but it comes
at an additional cost---we would have to accept the fact that
the universe is somewhat older (by a few billion years) than we
now believe. Actually, this situation is unavoidable for any cosmology
with $\omega<-1/3$ because of the relation between $R_0$ and
$ct_0$ in equation (8).

\vskip -0.1in
\begin{figure}
\center{\includegraphics[scale=0.5,angle=-90]{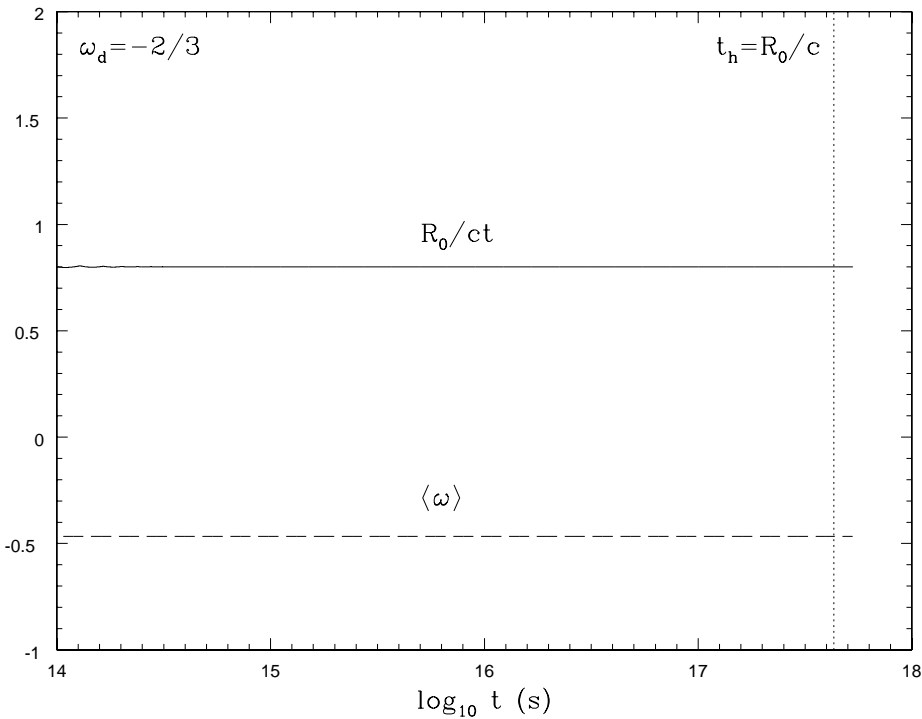}
\caption{Same as figure~1, except for a ``scaling solution"
in which $\rho_d\propto\rho_m$ and $\omega_d=-2/3$. The time-averaged equation-of-state
parameter is $\langle\omega\rangle\approx -0.47$. Thus, $R_0\not= ct_0$. Instead,
$t_0\approx 16.9$ billion years, approximately $3.4$ billion years longer than the
``horizon" time, $t_h\equiv R_0/c$ ($\approx 13.5$ billion years). This type of
universe is not subject to the ``coincidence" problem since $R_0/ct$ is constant.
It provides the best fit to the Type Ia supernova data (see figure~9).}}
\end{figure}
\vskip 0.1in

We see in figure~7 that $R_0/ct$ is constant, but at a value
$(3/2)(1+\langle\omega\rangle)$, where the time-averaged $\omega$
is now $\approx -0.47$. Thus, if $R_0$ is 13.5 billion light-years,
$t_0$ must be approximately 16.9 billion years. The explanation for
this (Melia 2008) is that what we believe to be the Universe's age
is actually the horizon time $t_h\equiv R_0/c$, which is shorter
than its actual age $t_0$. In figure~7, the distinction between
$t_h$ and $t_0$ is manifested primarily through the termination
points of the $R_0$ and $\langle\omega\rangle$ curves, which
extend past the vertical dotted line at $t=t_h$.

Ironically, this unexpected result has several important consequences,
such as offering an explanation for the early appearance of supermassive
black holes (at a redshift $>6$; see Fan et al. 2003), and the glaring
deficit of dwarf halos in the local group (see Klypin et al. 1999).
Both of these long-standing problems in cosmology would be resolved if
the Universe were older. Supermassive black holes would have had much
more time ($4-5$ billion years) to form than current thinking allows
(i.e., only $\sim 800$ million years), and dwarf halos would
correspondingly have had more time to merge hierarchically, depleting
the lower mass end of the distribution.

\vskip -0.1in
\begin{figure}
\center{\includegraphics[scale=0.5,angle=-90]{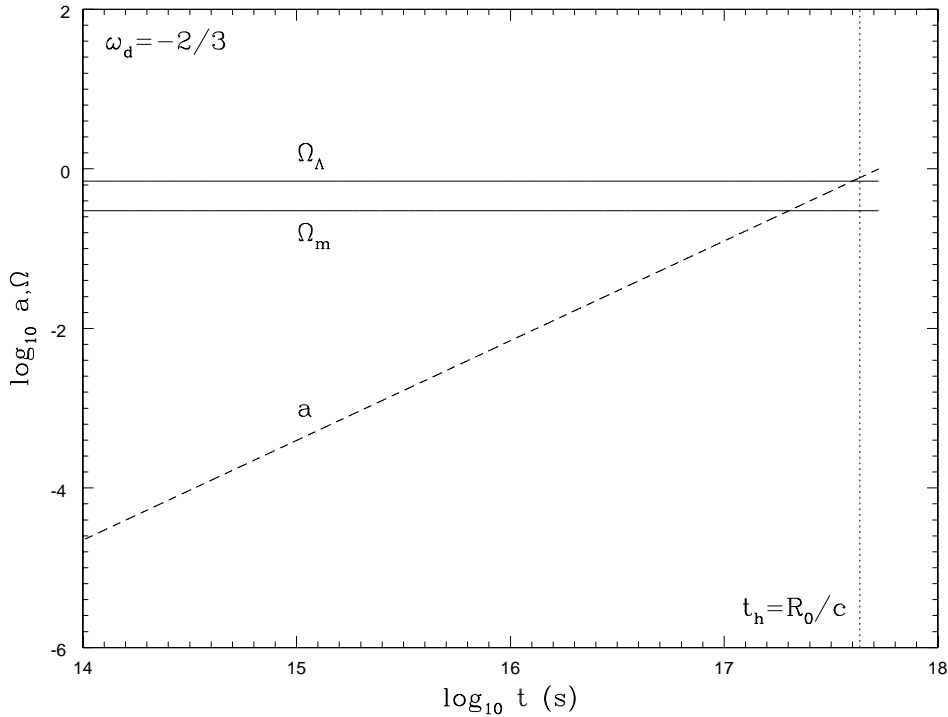}
\caption{Same as figure~2, except for a ``scaling solution"
in which $\rho_d\propto\rho_m$ and $\omega_d=-2/3$.}}
\end{figure}
\vskip 0.1in

The matter and dark energy densities corresponding to the $\omega_d=-2/3$
scaling solution are shown as functions of cosmic time in figure~8, along
with the evolution of the scale factor $a(t)$. Here too, the ``coincidence
problem" does not exist, and the flatness problem is resolved since
$kR_0^2/a^2\rightarrow 0$ as $t\rightarrow\infty$ in equation (10), so
the constant in equation (13) should be $\approx 1$ at late times,
regardless of the value of $k$. Very importantly, this model fits the
Type Ia supernova data very well, as shown in figure~9. The best fit
corresponds to $\Delta=25.26$, with a reduced $\chi^2=1.001$ for
$180-1=179$ d.o.f.

\begin{figure}
\center{\includegraphics[scale=0.5,angle=-90]{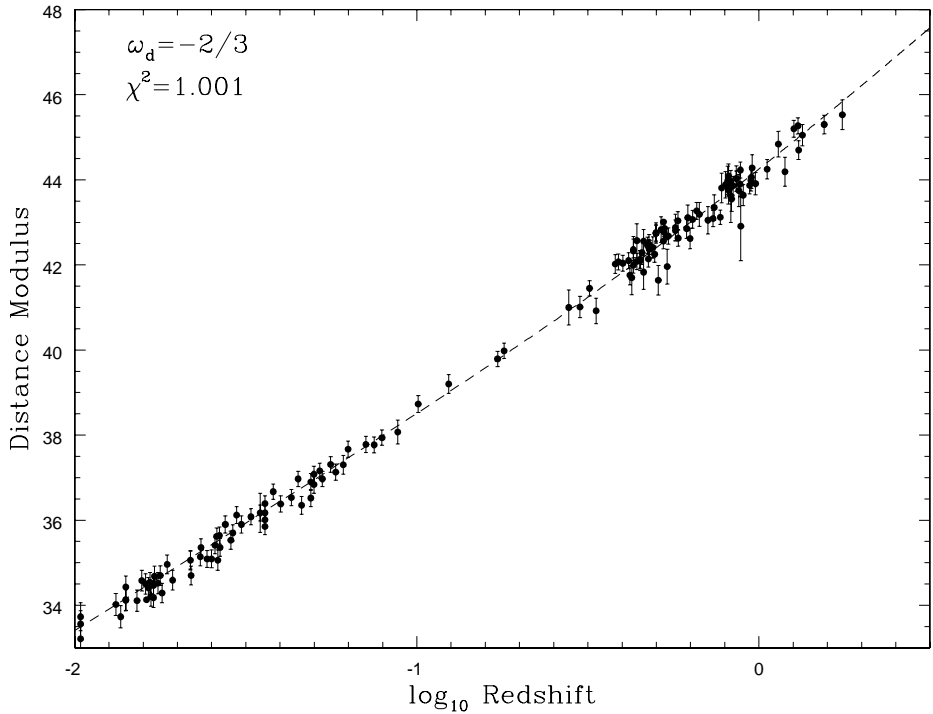}
\caption{Same as figure~3, except for a ``scaling solution"
in which $\rho_d\propto\rho_m$ and $\omega_d=-2/3$. This type of universe
provides the best fit to the Type Ia supernova data (Riess et al. 2004),
with a reduced $\chi^2\approx 1.001$ for $180-1=179$ d.o.f. The best fit
corresponds to $\Delta=25.26$.}}
\end{figure}

\section{Concluding Remarks}
Our main goal in this study has been to examine what we can learn
about the nature of dark energy from a consideration of the cosmic
horizon $R_0$ and its evolution with cosmic time. A principal
outcome of this work is the realization that the so-called
``coincidence" problem in the standard model is actually more
severe than previously thought. We have found that in a
$\Lambda$CDM universe, $R_0\rightarrow ct_0$ only once, and
according to the observations, this must be happening right now.
Coupled to the fact that the basic $\Lambda$CDM model does not
adequately account for the Type Ia supernova data without
introducing new parameters and patching together phases of
deceleration and acceleration that must somehow blend together
only in the current epoch, this is a strong indication that dark
energy almost certainly is not due to a cosmological constant.
Other issues that have already been discussed extensively in
the literature, such as the fact that the vacuum energy in
quantum theory should greatly exceed the required value of
$\Lambda$, only make this argument even more compelling. Of
course, this rejection of the cosmological-constant hypothesis
then intensifies interest into the question of why we don't
see any vacuum energy at all, but this is beyond the scope
of the present paper.

Many alternatives to a cosmological constant have been
proposed over the past decade but, for the sake of simplicity,
we have chosen in this paper to focus our attention on scaling
solutions. The existence of such cosmologies has been discussed
extensively in the literature, within the context of standard
General Relativity, braneworlds (Randall-Sundrum and Gauss-Bonnett),
and Cardassian scenarios, among others (see, e.g., Maeda 2001;
Freese and Lewis 2002; and Fujimoto and He 2004).

Our study has shown that scaling solutions not only fit the
Type Ia supernova data much better than the basic $\Lambda$CDM
cosmology, but they apparently simultaneously solve several
conundrums with the standard model. As long as the time-averaged
value of $\omega$ is less than $-1/3$, they eliminate both the
coincidence and flatness problems, possibly even obviating the
need for a period of rapid inflation in the early universe
(see, e.g., Guth 1981; Linde 1982).

But most importantly, as far as this study is concerned, scaling
solutions account very well for the observed fact that $R_0\approx 
ct_0$. If $\langle\omega\rangle=-1/3$ exactly, then $R_0(t)=ct$
for all cosmic time, and therefore the fact that we see this
condition in the present Universe is no coincidence at all.
On the other hand, if $\langle\omega\rangle<-1/3$, scaling
solutions fit the Type Ia supernova data even better, but
then we have to accept the conclusion that the Universe is
older than the horizon time $t_h\equiv R_0/c$. According to
our calculations, which produce a best fit to the supernova
data for $\langle\omega\rangle=-0.47$ (corresponding to a
dark energy equation of state with $\omega_d=-2/3$), the age
of the Universe should then be $t_0\approx 16.9$ billion years.
This may be surprising at first, perhaps even unbelievable to
some, but the fact of the matter is that such an age actually
solves other major problems in cosmology, including the (too)
early appearance of supermassive black holes, and the glaring
deficit of dwarf halos in the local group of galaxies.

When thinking about a dynamical dark energy, it is worth recalling
that scalar fields arise frequently in particle physics, including
string theory, and any of these may be appropriate candidates for
this mysterious new component of $\rho$. Actually, though we have
restricted our discussion to equations of state with $\omega_d\ge -1$,
it may even turn out that a dark energy with $\omega_d<-1$ is
providing the Universe's acceleration. Such a field is usually
referred to as a Phantom or a ghost. The simplest explanation
for this form of dark energy would be a scalar field with a
negative kinetic energy (Caldwell 2002). However, Phantom
fields are plagued by severe quantum instabilities, since
their energy density is unbounded from below and the vacuum
may acquire normal, positive energy fields (Carroll et al.
2003). We have therefore not included theories with $\omega_d<-1$
in our analysis here, though a further consideration of their
viability may be warranted as the data continue to improve.

On the observational front, the prospects for confirming or
rejecting some of the ideas presented in this paper look
very promising indeed. An eagerly anticipated mission, SNAP
(Rhodes et al. 2004), will constrain the nature of dark energy
in two ways. First, it will observe deeper Type Ia supernovae.
Second, it will attempt to use weak gravitational lensing
to probe foreground mass structures. If selected, SNAP should
be launched by 2020. An already funded mission, the Planck
CMB satellite, probably won't have the sensitivity to measure
any evolution in $\omega_d$, but it may be able to tell us
whether or not $\omega_d=-1$. An ESA mission, Planck is
scheduled for launch in mid-2008.

Finally, we may be on the verge of uncovering a class
of sources other than Type Ia supernovae to use for
dark-energy exploration. Type Ia supernovae have greatly
enhanced our ability to study the Universe's expansion out
to a redshift $\sim 2$. But this new class of sources may
possibly extend this range to values as high as $5-10$.
According to Hooper and Dodelson (2007), Gamma Ray Bursts
(GRBs) have the potential to detect dark energy with a
reasonable significance, particularly if there was an
appreciable amount of it at early times, as suggested by
scaling solutions. It is still too early to tell if GRBs
are good standard candles, but since differences between
$\Lambda$CDM and dynamical dark energy scenarios are more
pronounced at early times (see figures~3, 6, and 9),
GRBs may in the long run turn out to be even more important
than Type Ia supernovae in helping us learn about the true
nature of this unexpected ``third" form of energy.

\section*{Acknowledgments}

This research was partially supported by NSF grant 0402502 at the
University of Arizona. Many inspirational discussions with Roy Kerr
are greatly appreciated. Part of this work was carried out at
Melbourne University and at the Center for Particle Astrophysics
and Cosmology in Paris.

\end{document}